\documentclass[aps,twocolumn,prl,groupedaddress]{revtex4}
\usepackage{graphicx}
\usepackage{amssymb}
\begin{document}

\title{Morphology Scaling of Drop Impact onto a Granular Layer} 

\author{Hiroaki Katsuragi}
\affiliation{Department of Applied Science for Electronics and Materials, Kyushu University, 6-1 Kasugakoen, Kasuga, Fukuoka 816-8580, Japan}

\begin{abstract}
We investigate the impact of a free-falling water drop onto a granular layer. First, we constructed a phase diagram of crater shapes with two control parameters, impact speed and grain size. A low-speed impact makes a deeper cylindrical crater in a fluffy granular target. After high-speed impacts, we observed a convex bump higher than the initial surface level instead of a crater. The inner ring can be also observed in medium impact speed regime. Quantitatively, we found a scaling law for crater radius with a dimensionless number consisting of impact speed and density ratio between the bulk granular layer and water drop. This scaling demonstrates that the water drop deformation is crucial to understand the crater morphology.
\end{abstract}

\pacs{45.70.-n, 45.70.Mg, 47.57.Gc, 83.80.Fg}

\date{\today}

\maketitle
What happens when a drop of liquid impacts a granular layer? Despite recent developments in the fundamental physics of granular systems and fluid dynamics, this simple question remains unanswered. Much attention has been paid to dry granular materials~\cite{Duran2000}. The mechanical properties of wet cohesive granular matter have been almost the only aspect examined so far~\cite{Albert1997,Mitarai2006}. Recently, solid projectile impacts onto a granular layer have been well studied~\cite{Amato1998,Thoroddsen2001,Uehara2003,Walsh2003,Lohse2004N,Lohse2004P,Royer2005,deVet2007,Katsuragi2007,Nelson2008,Goldman2008}. In contrast, impacts between a drop and a hard wall or a fluid pool have been examined extensively~\cite{Yarin2006,Range1998,Marmanis1996}. Drop impact dynamics is related to many industrial applications, e.g., ink-jet printing, rapid spray cooling, and surface coating.  
Nevertheless, a drop impact to a granular layer has not been investigated until quite recently~\cite{Delon2009}.
Here, we focus on the drop-granular impact. Our simple experiments serve as a starting point for exploring this phenomenon, which has great potential applicability to various fields such as planetary science, material science, civil engineering, and agriculture. 
For instance,  the drop-granular impact may help to understand a geological scale impact in which a projectile is destroyed completely by the impact~\cite{Nagao2005}. It may also relate to {\it the fossil rain drops} which are small circular pit-like depressions in fine grain sediment and whose origin is still a subject of controversy~\cite{Desor1850,Metz1981}.

We perform simple granular impact experiments with a free-falling water drop. When the drop impacts a granular layer, it rebounds from the surface and then slowly sinks into the granular layer. Afterwards, a crater remains as evidence of the impact. Impact speed $v$ is controlled by free-fall height $h$~\cite{neglect1}, which ranges in this experiment from $10$ to $480$~mm. The granular grains, which are commercial SiC abrasives, possess nonspherical shapes and polydispersity and their grain size $D_g$ is varied from $4$, $8$, $14$, $20$, or $50$~$\mu$m. A small vessel ($30$ mm in diameter, $10$ mm thick) is filled with grains by hand and used as a target. The five grain sizes can be grouped into three classes in terms of the packing fraction: $0.31$ ($D_g=4,8$~$\mu$m), $0.44$ ($D_g=14,20$~$\mu$m), and $0.50$ ($D_g=50$~$\mu$m). The uncertainty of the packing fraction is about $10\%$. These packing fraction values are much smaller than that of random close packing. Thus, the granular layer includes numerous pores. Such a low packing fraction in $10^0$-$\mu$m-size grains results from their irregular shape and polydispersity~\cite{Roller1930,Suzuki2001}. The radius of a drop is fixed at $R_w =2.4 \pm 0.2$~mm. 

\begin{figure*}
\begin{center}
\scalebox{0.92}[0.92]{\includegraphics{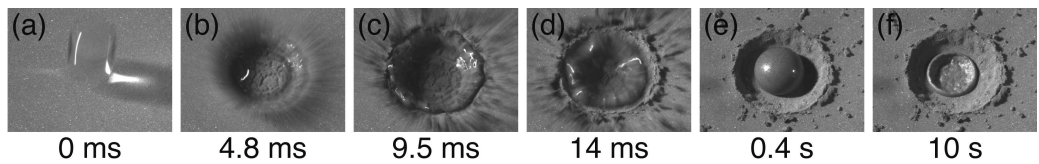}}
\caption{Typical sequence of a drop-granular impact ($D_g=4$~$\mu$m and $h=160$~mm). The transition from drop expansion to receding occurs between (c) and (d). Drop deformation creates a crater rim at this early stage. The slow sinking of the drop forms an inner ring at a later stage~(e,f). 
}
\label{fig:imgs}
\end{center}
\end{figure*}

Figure~\ref{fig:imgs} shows a typical sequence of a drop-granular impact taken by a high-speed camera (TAKEX FC350CL) at $210$~fps. The first four images~(Fig.~\ref{fig:imgs}(a-d)) display the impact moment. We observe the great deformation of the drop, which forms the crater rim. This rim forms within approximately $10$~ms. After this initial impact stage, the drop rebounds and undergoes attenuated oscillation for a while. Then, the drop remains still on the surface as shown in Fig.~\ref{fig:imgs}(e). The oscillation settling time is $0.4$~s. Finally, the drop penetrates very slowly into the granular layer and then forms a ring shape inside the crater rim (Fig.~\ref{fig:imgs}(f)); this sinking process takes $10$~s. The time scales of early deformation and the later sinking are $10^{-2}$~s and $10^1$~s, respectively. Thus, the time scale expands by about three orders of magnitude. 

From an inspection of all video data, we find that the first expansive drop deformation time scale $t_d$ does not depend on $D_g$ and $h$. It is always $t_d \sim 10^{-2}$~s. This time scale seems to be determined purely by the properties of the water drop. The surface tension of the water drop, $\gamma=7.2 \times 10^{-2}$~N/m and its mass $m_w=5.8\times 10^{-5}$~kg yield the time scale $t_{\gamma}=\sqrt{m_w / \gamma} = 2.8 \times 10^{-2}$~s. $t_{\gamma}$ should be the period of drop oscillation~\cite{Okumura2003,Biance2006}. This estimate is consistent with the experimental result, i.e., $t_{\gamma} \sim 2 t_{d}$. 

\begin{figure}
\begin{center}
\scalebox{0.92}[0.92]{\includegraphics{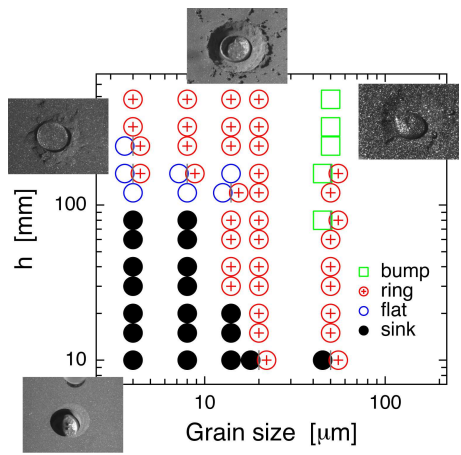}}
\caption{(color online). Phase diagram of crater shapes resulting from drop-granular impact. Twelve free-fall heights and five grain sizes are used in the  experiments. In each experimental condition, two independent runs are conducted; i.e., $120$ impacts are examined in total. The presence of two symbols at a point indicates that two types are possible at that point. We have four types of crater shapes depending on $D_g$ and $h$; sink, ring, flat, and bump. The characteristic $h$ at which the qualitative crater shape changes is $100$~mm.
}
\label{fig:pd}
\end{center}
\end{figure}

The shape of the crater is not limited to that shown in Fig.~\ref{fig:imgs}. It depends on the impact speed $v$ and grain size $D_g$. We classify the crater shapes into four types, and draw a phase diagram of crater shapes~(Fig.~\ref{fig:pd}). At low impact speed, clear splashing cannot be observed, and the drop sinks very gradually, leaving a cylindrical crater. It seems that the penetrating water drop compresses the fluffy granular layer mainly by capillary effect. We call this a ``sink crater''. 
As the impact speed increases, the splashing of grains begins to emerge, and a ``ring crater'' is created~(Fig.~\ref{fig:imgs}). Between these two cases, a ``flat crater'' can be observed in the case of relatively small grains. In this type, the global structure of the crater is similar to that of the ring type, but the central region is rather flat. A ``bump crater'' appears at a high impact speed and large grain size. In this regime, the drop splits into smaller parts upon impact, and the largest part remains at the center. It absorbs surrounding surface grains at the impact stage, and forms a convex bump by the sedimentation of collected grains during its penetration. Thus, drop impact can create even a convex shape as well as concave craters. 

\begin{figure*}
\begin{center}
\scalebox{0.92}[0.92]{\includegraphics{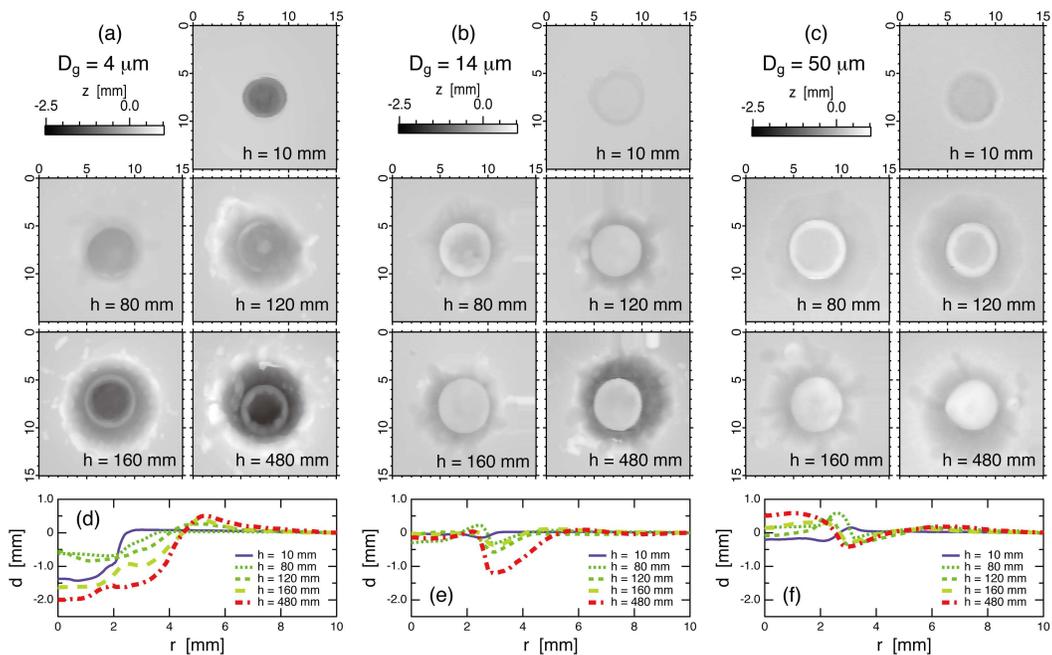}}
\caption{(color online). Grayscale surface depth maps of $D_g=$ (a)~$4$, (b)~$14$, and (c)~$50$~$\mu$m cases. The vertical height $z=0$ is the initial granular surface level before impact. All depth map data are shown in $15 \times 15$ mm$^2$ squares with axes in units of mm. The corresponding radial depth $d$ as a function of distance $r$ from the center of the crater is shown in (d-f). As (a,d) show, $h=10$~mm results in deeper craters than $h=80$ or $120$~mm cases. In large $D_g$ impacts, it is hard to observe a hemispherical crater. Furthermore, the central bump is higher than the initial height $z=0$ when $D_g=50$~$\mu$m and $h=480$~mm. This is called a bump crater in the phase diagram shown in Fig.~\ref{fig:pd}.}
\label{fig:surfs}
\end{center}
\end{figure*}

To characterize the crater shape precisely, its surface structure is measured by a line laser displacement sensor system (KEYENCE LJG030)~\cite{Mizoue2010}. 
Typical results are shown in Fig.~\ref{fig:surfs}. According to the phase diagram, $h=100$~mm is a characteristic free-fall height that corresponds roughly to the boundary of the sink, flat, and bump types. Therefore, we show the cases of $h=80,120,$ and $160$~mm in Fig.~\ref{fig:surfs}. We also show $h=10$ and $480$~mm, wchich correspond to the slowest and the fastest impacts, respectively. 

From the surface depth map data~(Fig.~\ref{fig:surfs}(a-c)), we compute the radial depth function $d(r)$, where $r$ is the distance from the center of the crater, and plot the corresponding $d(r)$ curves~(Fig.~\ref{fig:surfs}(d-f)). When $D_g=4$~$\mu$m, the shortest height ($h=10$~mm) impact creates a deeper crater than medium-height impacts of $h=80$ and $120$~mm. 
In medium-height impacts, the impact inertia is inadequate to make a deep crater, but it compresses the granular layer a little. And it effectively suppresses the capillary based compression. Therefore a lower impact energy is better able to deform a fluffy granular layer. 
The crater depth is not a monotonic function; when the impact speed is large enough, craters become deeper again~(Fig.~\ref{fig:surfs}(a,d)). With large grains, the crater shapes are completely different as shown in Fig.~\ref{fig:surfs}(b,c,e,f). They are no longer simple concave shapes as mentioned before. 
Interestingly, the bump height is much higher than the initial granular layer level $z=0$~(Fig.~\ref{fig:surfs}(f), for $h=480$~mm).

A solid projectile impact cratering was also measured using a laser profilometry~\cite{deVet2007}. It was found that the crater shapes are very close to hyperbolic function. However, the drop-granular impact creates very different craters due to the drop deformation and capillary effect, as mentioned above. 

\begin{figure}
\begin{center}
\scalebox{0.92}[0.92]{\includegraphics{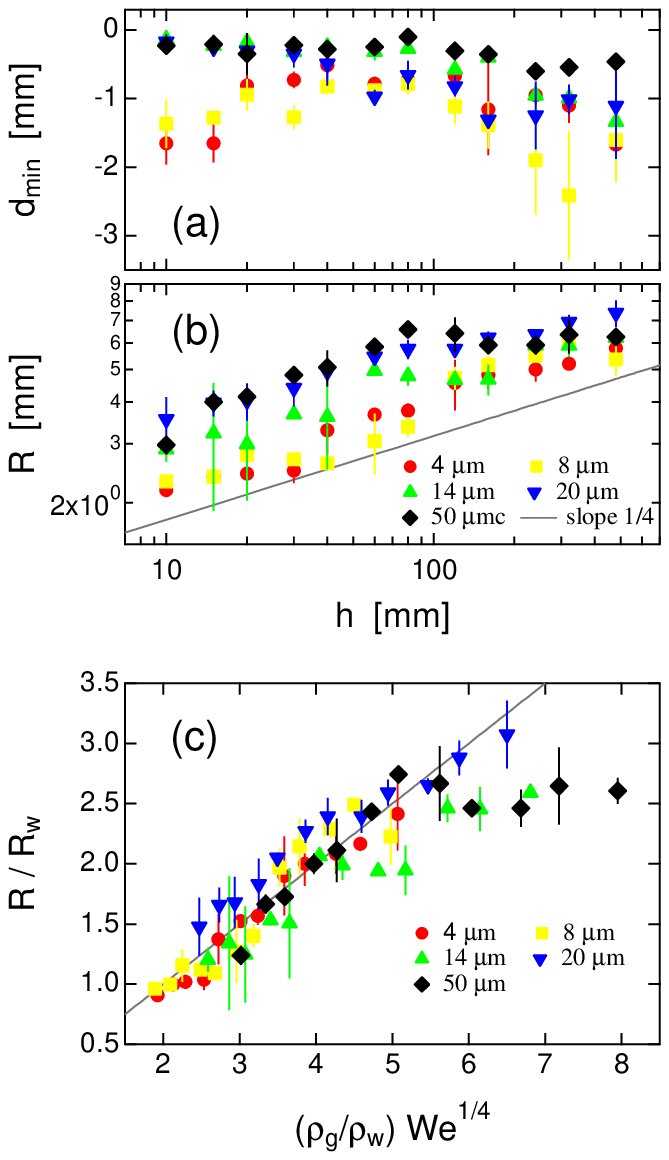}}
\caption{(color online). Characteristic length scales. (a)~Minimum crater depth $d_{min}$ and (b)~crater radius $R$, as functions of free-fall height $h$. $d_{min}$ shows a peak at $h \sim 100$~mm at $D_g=4$ and $8$~$\mu$m cases~(a). Other curves are slightly decreasing functions of $h$. $R$ roughly agrees with the scaling $R \sim h^{1/4}$~(b). To collapse all the $R$ data, a dimensionless plot $R/R_w$ vs $(\rho_g/\rho_w)We^{1/4}$ is shown~(c). We can confirm good data collapse to the scaling. In all plots, symbols and colors indicate grain size $D_g$ of the granular layer, as shown in the legend.
}
\label{fig:scaling}
\end{center}
\end{figure}

Next, we analyze the characteristic length scales. First, we discuss the vertical length scale. The minimum crater depth $d_{min}$ with respect to free-fall height $h$ is shown in Fig.~\ref{fig:scaling}(a). The $d_{min}$ curves for small $D_g$ have a peak at a certain $h$~($\sim 100$~mm). On the other hand, the $d_{min}$ curves for large $D_g$ are rather weakly decreasing functions of $h$. 

Analysis of the horizontal length scale $R$, the radius of the crater, is more straightforward. Here, we assume $R$ is determined by the deformed drop radius $R_d$. This assumption is roughly confirmed in Fig.~\ref{fig:imgs} (c,d). Water drop deformation by the impact is estimated by Okumura et al.~\cite{Okumura2003,Biance2006}. They studied water drop impact onto a super-hydrophobic substrate~\cite{Richard2000}. Using Euler's equation, the Laplace pressure gradient, and geometrical conditions, they have derived a horizontal deformation $R_d$ that is scaled as $R_d \sim R_w We^{1/4}$, where $We$ is the Weber number, $We= 2 \rho_w R_w v^2 / \gamma$, and $\rho_w$ is the density of water. $We$ is proportional to $h$ through the free-fall relation $v^2 \sim h$. Thus, $R$ should scale as $R \sim h^{1/4}$. The actual $R$ data and this scaling slope are shown in Fig.~\ref{fig:scaling}(b). Although the global structure seems to be sound, the data spread depending on $D_g$. We believe this spreading is caused by the difference between the density of the bulk granular layer $\rho_g$ and that of the water drop $\rho_w$. Of course, $\rho_g$ is essentially the same parameter as  the packing fraction. If the ratio $\rho_g/\rho_w$ is large, the drop is greatly deformed by the impact. Consequently, a large crater is created. The $\rho_g$ is the simplest representative property of bulk granular systems. Thus, it is natural to consider the bulk density ratio when discussing the granular impact physics. In fact, this bulk density ratio is also useful for discussing the dynamic scaling of solid projectile impacts to a granular layer~\cite{Katsuragi2010}. Using this density ratio, we finally define the scaling as, 
\begin{equation}
\frac{R}{R_w}\sim \frac{\rho_g}{\rho_w}We^{1/4}.
\label{eq:scaling}
\end{equation}
To check this scaling, the normalized crater radius $R/R_w$ is plotted as a function of the scaling parameter $(\rho_g/\rho_w)We^{1/4}$~(Fig.~\ref{fig:scaling}(c)). This scaling implies $R \sim \rho_g ( R_w^5 v^2 / \rho_w^3 \gamma )^{1/4}$. As expected, the data collapse to a line. Although the original scaling $R_d \sim We^{1/4}$ was deduced for a hard wall target, it is still valid for a deformable granular target. At large $We$, data deviate from the scaling due to the drop splitting by the high impact energy.

Similar crater radius scaling was found in solid projectile impact experiments~\cite{Amato1998,Uehara2003,Walsh2003}. The scaling exponent $1/4$ means that the most of the impact energy is spent in lifting the ejecta resulting from cratering~\cite{Amato1998}. If this is true, the crater size should be smaller with larger $\rho_g$. However, the current result shows the opposite tendency. 
The physics of solid impacts and drop impacts are quite different, even though the crater radius scaling exponent is similar. 
In drop-granular impact, drop deformation effect is much more crucial than ejecta splashing. The impact Reynolds number $I=We^{1/4}Re^{1/2}$ is derived by substituting the water drop characteristic length scale and time scale into the definition of the Reynolds number $Re=v\sqrt{\nu t_{\gamma}}/\nu$ with kinematic viscosity $\nu$~\cite{Marmanis1996}. A systematic evaluation of the impacting fluid's surface tension and viscosity dependency is necessary for using an advanced dimensionless number such as $I$. This is an open problem. 

We have demonstrated a simple but original experiment of drop-granular impact. The drop deformation time scale can be evaluated by the surface-tension-based elasticity of the water drop. We classified the resulting crater shapes into four characteristic types and completed the phase diagram. We found that the crater radius scales as $R/R_w \sim (\rho_g/\rho_w)We^{1/4}$. However, the vertical length behaves in a more complex manner. It seems to be affected by competition between multiple fluid-granular interactions. To reveal the effect more accurately and obtain more universal scaling, we must systematically vary many parameters, e.g., viscosity, surface tension, size of impacting drop, and grain shape and surface properties. This report is the first step toward revealing this rich drop-granular impact world. 

\begin{acknowledgments}
The author thanks K. Okumura for useful discussion. This research has been supported by the Japanese Ministry of Education, Culture, Sports, Science and Technology (MEXT), Grant-in-Aid for Young Scientists, No.~21684021.
\end{acknowledgments}

\bibliography{dropimpact}

\begin{thebibliography}{29}
\expandafter\ifx\csname natexlab\endcsname\relax\def\natexlab#1{#1}\fi
\expandafter\ifx\csname bibnamefont\endcsname\relax
  \def\bibnamefont#1{#1}\fi
\expandafter\ifx\csname bibfnamefont\endcsname\relax
  \def\bibfnamefont#1{#1}\fi
\expandafter\ifx\csname citenamefont\endcsname\relax
  \def\citenamefont#1{#1}\fi
\expandafter\ifx\csname url\endcsname\relax
  \def\url#1{\texttt{#1}}\fi
\expandafter\ifx\csname urlprefix\endcsname\relax\def\urlprefix{URL }\fi
\providecommand{\bibinfo}[2]{#2}
\providecommand{\eprint}[2][]{\url{#2}}

\bibitem[{\citenamefont{Duran}(2000)}]{Duran2000}
\bibinfo{author}{\bibfnamefont{J.}~\bibnamefont{Duran}},
  \emph{\bibinfo{title}{Sands, powders, and grains: An introduction to the
  physics of granular materials}} (\bibinfo{publisher}{Springer},
  \bibinfo{address}{New York}, \bibinfo{year}{2000}).

\bibitem[{\citenamefont{Albert et~al.}(1997)\citenamefont{Albert, Albert,
  Hornbaker, Schiffer, and Barab{\'a}si}}]{Albert1997}
\bibinfo{author}{\bibfnamefont{R.}~\bibnamefont{Albert}},
  \bibinfo{author}{\bibfnamefont{I.}~\bibnamefont{Albert}},
  \bibinfo{author}{\bibfnamefont{D.}~\bibnamefont{Hornbaker}},
  \bibinfo{author}{\bibfnamefont{P.}~\bibnamefont{Schiffer}}, \bibnamefont{and}
  \bibinfo{author}{\bibfnamefont{A.~L.} \bibnamefont{Barab{\'a}si}},
  \bibinfo{journal}{Phys. Rev. E} \textbf{\bibinfo{volume}{56}},
  \bibinfo{pages}{R6271} (\bibinfo{year}{1997}).

\bibitem[{\citenamefont{Mitarai and Nori}(2006)}]{Mitarai2006}
\bibinfo{author}{\bibfnamefont{N.}~\bibnamefont{Mitarai}} \bibnamefont{and}
  \bibinfo{author}{\bibfnamefont{F.}~\bibnamefont{Nori}},
  \bibinfo{journal}{Adv. Phys.} \textbf{\bibinfo{volume}{55}},
  \bibinfo{pages}{1} (\bibinfo{year}{2006}).

\bibitem[{\citenamefont{Amato and Williams}(1998)}]{Amato1998}
\bibinfo{author}{\bibfnamefont{J.~C.} \bibnamefont{Amato}} \bibnamefont{and}
  \bibinfo{author}{\bibfnamefont{R.~E.} \bibnamefont{Williams}},
  \bibinfo{journal}{Am. J. Phys.} \textbf{\bibinfo{volume}{66}},
  \bibinfo{pages}{141} (\bibinfo{year}{1998}).

\bibitem[{\citenamefont{Thoroddsen and Shen}(2001)}]{Thoroddsen2001}
\bibinfo{author}{\bibfnamefont{S.~T.} \bibnamefont{Thoroddsen}}
  \bibnamefont{and} \bibinfo{author}{\bibfnamefont{A.~Q.} \bibnamefont{Shen}},
  \bibinfo{journal}{Phys. Fluids} \textbf{\bibinfo{volume}{13}},
  \bibinfo{pages}{4} (\bibinfo{year}{2001}).

\bibitem[{\citenamefont{Uehara et~al.}(2003)\citenamefont{Uehara, Ambrosso,
  Ojha, and Durian}}]{Uehara2003}
\bibinfo{author}{\bibfnamefont{J.~S.} \bibnamefont{Uehara}},
  \bibinfo{author}{\bibfnamefont{M.~A.} \bibnamefont{Ambrosso}},
  \bibinfo{author}{\bibfnamefont{R.~P.} \bibnamefont{Ojha}}, \bibnamefont{and}
  \bibinfo{author}{\bibfnamefont{D.~J.} \bibnamefont{Durian}},
  \bibinfo{journal}{Phys. Rev. Lett.} \textbf{\bibinfo{volume}{90}},
  \bibinfo{pages}{194301} (\bibinfo{year}{2003}).

\bibitem[{\citenamefont{Walsh et~al.}(2003)\citenamefont{Walsh, Holloway,
  Habdas, and de~Bruyn}}]{Walsh2003}
\bibinfo{author}{\bibfnamefont{A.~M.} \bibnamefont{Walsh}},
  \bibinfo{author}{\bibfnamefont{K.~E.} \bibnamefont{Holloway}},
  \bibinfo{author}{\bibfnamefont{P.}~\bibnamefont{Habdas}}, \bibnamefont{and}
  \bibinfo{author}{\bibfnamefont{J.~R.} \bibnamefont{de~Bruyn}},
  \bibinfo{journal}{Phys. Rev. Lett.} \textbf{\bibinfo{volume}{91}},
  \bibinfo{pages}{104301} (\bibinfo{year}{2003}).

\bibitem[{\citenamefont{Lohse et~al.}(2001)\citenamefont{Lohse, Rauh{\'e},
  Bergmann, and van~der Meer}}]{Lohse2004N}
\bibinfo{author}{\bibfnamefont{D.}~\bibnamefont{Lohse}},
  \bibinfo{author}{\bibfnamefont{R.}~\bibnamefont{Rauh{\'e}}},
  \bibinfo{author}{\bibfnamefont{R.}~\bibnamefont{Bergmann}}, \bibnamefont{and}
  \bibinfo{author}{\bibfnamefont{D.}~\bibnamefont{van~der Meer}},
  \bibinfo{journal}{Nature} \textbf{\bibinfo{volume}{432}},
  \bibinfo{pages}{689} (\bibinfo{year}{2001}).

\bibitem[{\citenamefont{Lohse et~al.}(2004)\citenamefont{Lohse, Bergmann,
  Mikkelsen, Zeilstra, van~der Meer, Versluis, van~der Weele, van~der Hoef, and
  Kuipers}}]{Lohse2004P}
\bibinfo{author}{\bibfnamefont{D.}~\bibnamefont{Lohse}},
  \bibinfo{author}{\bibfnamefont{R.}~\bibnamefont{Bergmann}},
  \bibinfo{author}{\bibfnamefont{R.}~\bibnamefont{Mikkelsen}},
  \bibinfo{author}{\bibfnamefont{C.}~\bibnamefont{Zeilstra}},
  \bibinfo{author}{\bibfnamefont{D.}~\bibnamefont{van~der Meer}},
  \bibinfo{author}{\bibfnamefont{M.}~\bibnamefont{Versluis}},
  \bibinfo{author}{\bibfnamefont{K.}~\bibnamefont{van~der Weele}},
  \bibinfo{author}{\bibfnamefont{M.}~\bibnamefont{van~der Hoef}},
  \bibnamefont{and} \bibinfo{author}{\bibfnamefont{H.}~\bibnamefont{Kuipers}},
  \bibinfo{journal}{Phys. Rev. Lett.} \textbf{\bibinfo{volume}{93}},
  \bibinfo{pages}{198003} (\bibinfo{year}{2004}).

\bibitem[{\citenamefont{Royer et~al.}(2005)\citenamefont{Royer, Corwin, Flior,
  Cordero, Rivers, Eng, and Jaeger}}]{Royer2005}
\bibinfo{author}{\bibfnamefont{J.~R.} \bibnamefont{Royer}},
  \bibinfo{author}{\bibfnamefont{E.~I.} \bibnamefont{Corwin}},
  \bibinfo{author}{\bibfnamefont{A.}~\bibnamefont{Flior}},
  \bibinfo{author}{\bibfnamefont{M.-L.} \bibnamefont{Cordero}},
  \bibinfo{author}{\bibfnamefont{M.~L.} \bibnamefont{Rivers}},
  \bibinfo{author}{\bibfnamefont{P.~J.} \bibnamefont{Eng}}, \bibnamefont{and}
  \bibinfo{author}{\bibfnamefont{H.~M.} \bibnamefont{Jaeger}},
  \bibinfo{journal}{Nature Phys.} \textbf{\bibinfo{volume}{1}},
  \bibinfo{pages}{164} (\bibinfo{year}{2005}).

\bibitem[{\citenamefont{de~Vet and de~Bruyn}(2007)}]{deVet2007}
\bibinfo{author}{\bibfnamefont{S.~J.} \bibnamefont{de~Vet}} \bibnamefont{and}
  \bibinfo{author}{\bibfnamefont{J.~R.} \bibnamefont{de~Bruyn}},
  \bibinfo{journal}{Phys. Rev. E} \textbf{\bibinfo{volume}{76}},
  \bibinfo{pages}{041306} (\bibinfo{year}{2007}).

\bibitem[{\citenamefont{Katsuragi and Durian}(2007)}]{Katsuragi2007}
\bibinfo{author}{\bibfnamefont{H.}~\bibnamefont{Katsuragi}} \bibnamefont{and}
  \bibinfo{author}{\bibfnamefont{D.~J.} \bibnamefont{Durian}},
  \bibinfo{journal}{Nature Phys.} \textbf{\bibinfo{volume}{3}},
  \bibinfo{pages}{420} (\bibinfo{year}{2007}).

\bibitem[{\citenamefont{Nelson et~al.}(2008)\citenamefont{Nelson, Katsuragi,
  Mayor, and Durian}}]{Nelson2008}
\bibinfo{author}{\bibfnamefont{E.~L.} \bibnamefont{Nelson}},
  \bibinfo{author}{\bibfnamefont{H.}~\bibnamefont{Katsuragi}},
  \bibinfo{author}{\bibfnamefont{P.}~\bibnamefont{Mayor}}, \bibnamefont{and}
  \bibinfo{author}{\bibfnamefont{D.~J.} \bibnamefont{Durian}},
  \bibinfo{journal}{Phys. Rev. Lett.} \textbf{\bibinfo{volume}{101}},
  \bibinfo{pages}{068001} (\bibinfo{year}{2008}).

\bibitem[{\citenamefont{Goldman and Umbanhowar}(2008)}]{Goldman2008}
\bibinfo{author}{\bibfnamefont{D.~I.} \bibnamefont{Goldman}} \bibnamefont{and}
  \bibinfo{author}{\bibfnamefont{P.}~\bibnamefont{Umbanhowar}},
  \bibinfo{journal}{Phys. Rev. E} \textbf{\bibinfo{volume}{77}},
  \bibinfo{pages}{021308} (\bibinfo{year}{2008}).

\bibitem[{\citenamefont{Yarin}(2006)}]{Yarin2006}
\bibinfo{author}{\bibfnamefont{A.~L.} \bibnamefont{Yarin}},
  \bibinfo{journal}{Ann. Rev. Fluid Mech.} \textbf{\bibinfo{volume}{38}},
  \bibinfo{pages}{159} (\bibinfo{year}{2006}).

\bibitem[{\citenamefont{Range and Feuillebois}(1998)}]{Range1998}
\bibinfo{author}{\bibfnamefont{K.}~\bibnamefont{Range}} \bibnamefont{and}
  \bibinfo{author}{\bibfnamefont{F.}~\bibnamefont{Feuillebois}},
  \bibinfo{journal}{J. Colloid Interface Sci.} \textbf{\bibinfo{volume}{203}},
  \bibinfo{pages}{16} (\bibinfo{year}{1998}).

\bibitem[{\citenamefont{Marmanis and Thoroddsen}(1996)}]{Marmanis1996}
\bibinfo{author}{\bibfnamefont{H.}~\bibnamefont{Marmanis}} \bibnamefont{and}
  \bibinfo{author}{\bibfnamefont{S.~T.} \bibnamefont{Thoroddsen}},
  \bibinfo{journal}{Phys. Fluids} \textbf{\bibinfo{volume}{8}},
  \bibinfo{pages}{1344} (\bibinfo{year}{1996}).

\bibitem[{\citenamefont{Delon et~al.}(2009)\citenamefont{Delon, Dorbolo,
  Vandewalle, and Caps}}]{Delon2009}
\bibinfo{author}{\bibfnamefont{G.}~\bibnamefont{Delon}},
  \bibinfo{author}{\bibfnamefont{S.}~\bibnamefont{Dorbolo}},
  \bibinfo{author}{\bibfnamefont{N.}~\bibnamefont{Vandewalle}},
  \bibnamefont{and} \bibinfo{author}{\bibfnamefont{H.}~\bibnamefont{Caps}},
  \bibinfo{journal}{62nd Annual Meeting of the APS Division of Fluid Dynamics}
  \textbf{\bibinfo{volume}{54}}, \bibinfo{pages}{AH.00006}
  (\bibinfo{year}{2009}).

\bibitem[{\citenamefont{Nagao et~al.}(2005)\citenamefont{Nagao, Kibe, Shimizu,
  and Hikiji}}]{Nagao2005}
\bibinfo{author}{\bibfnamefont{Y.}~\bibnamefont{Nagao}},
  \bibinfo{author}{\bibfnamefont{S.}~\bibnamefont{Kibe}},
  \bibinfo{author}{\bibfnamefont{T.}~\bibnamefont{Shimizu}}, \bibnamefont{and}
  \bibinfo{author}{\bibfnamefont{M.}~\bibnamefont{Hikiji}},
  \bibinfo{journal}{56th International Astronautical Congress} pp.
  \bibinfo{pages}{IAC--05--B6.4.05} (\bibinfo{year}{2005}).

\bibitem[{\citenamefont{Desor}(1850)}]{Desor1850}
\bibinfo{author}{\bibfnamefont{E.}~\bibnamefont{Desor}},
  \bibinfo{journal}{Edinburgh New Phil. J.} \textbf{\bibinfo{volume}{49}},
  \bibinfo{pages}{246 } (\bibinfo{year}{1850}).

\bibitem[{\citenamefont{Metz}(1981)}]{Metz1981}
\bibinfo{author}{\bibfnamefont{R.}~\bibnamefont{Metz}}, \bibinfo{journal}{J.
  Sediment. Petrol.} \textbf{\bibinfo{volume}{51}}, \bibinfo{pages}{265 }
  (\bibinfo{year}{1981}).

\bibitem[{neg()}]{neglect1}
\bibinfo{note}{We assume the free-fall relation $v=\sqrt{2gh}$, where $g$ is
  the gravitational acceleration ($g=9.8$~m/s$^2$). We neglect the air drag
  deceleration; air drag is negligible for a free-fall drop of
  $h<500$~mm~\cite{Range1998}.}

\bibitem[{\citenamefont{Roller}(1930)}]{Roller1930}
\bibinfo{author}{\bibfnamefont{P.~S.} \bibnamefont{Roller}},
  \bibinfo{journal}{Ind. Eng. Chem.} \textbf{\bibinfo{volume}{22}},
  \bibinfo{pages}{1206} (\bibinfo{year}{1930}).

\bibitem[{\citenamefont{Suzuki et~al.}(2001)\citenamefont{Suzuki, Sato,
  Hasegawa, and Hirota}}]{Suzuki2001}
\bibinfo{author}{\bibfnamefont{M.}~\bibnamefont{Suzuki}},
  \bibinfo{author}{\bibfnamefont{H.}~\bibnamefont{Sato}},
  \bibinfo{author}{\bibfnamefont{M.}~\bibnamefont{Hasegawa}}, \bibnamefont{and}
  \bibinfo{author}{\bibfnamefont{M.}~\bibnamefont{Hirota}},
  \bibinfo{journal}{Powder Tech.} \textbf{\bibinfo{volume}{118}},
  \bibinfo{pages}{53} (\bibinfo{year}{2001}).

\bibitem[{\citenamefont{Okumura et~al.}(2003)\citenamefont{Okumura, Chevy,
  Richard, Qu{\'e}r{\'e}, and Clanet}}]{Okumura2003}
\bibinfo{author}{\bibfnamefont{K.}~\bibnamefont{Okumura}},
  \bibinfo{author}{\bibfnamefont{F.}~\bibnamefont{Chevy}},
  \bibinfo{author}{\bibfnamefont{D.}~\bibnamefont{Richard}},
  \bibinfo{author}{\bibfnamefont{D.}~\bibnamefont{Qu{\'e}r{\'e}}},
  \bibnamefont{and} \bibinfo{author}{\bibfnamefont{C.}~\bibnamefont{Clanet}},
  \bibinfo{journal}{Europhys. Lett.} \textbf{\bibinfo{volume}{62}},
  \bibinfo{pages}{237} (\bibinfo{year}{2003}).

\bibitem[{\citenamefont{Biance et~al.}(2006)\citenamefont{Biance, Chevy,
  Clanet, Lagubeau, and Qu{\'e}r{\'e}}}]{Biance2006}
\bibinfo{author}{\bibfnamefont{A.-L.} \bibnamefont{Biance}},
  \bibinfo{author}{\bibfnamefont{F.}~\bibnamefont{Chevy}},
  \bibinfo{author}{\bibfnamefont{C.}~\bibnamefont{Clanet}},
  \bibinfo{author}{\bibfnamefont{G.}~\bibnamefont{Lagubeau}}, \bibnamefont{and}
  \bibinfo{author}{\bibfnamefont{D.}~\bibnamefont{Qu{\'e}r{\'e}}},
  \bibinfo{journal}{J. Fluid. Mech.} \textbf{\bibinfo{volume}{554}},
  \bibinfo{pages}{47} (\bibinfo{year}{2006}).

\bibitem[{\citenamefont{Mizoue et~al.}()\citenamefont{Mizoue, Aoki, Tokita,
  Honjo, Barraza, and Katsuragi}}]{Mizoue2010}
\bibinfo{author}{\bibfnamefont{T.}~\bibnamefont{Mizoue}},
  \bibinfo{author}{\bibfnamefont{Y.}~\bibnamefont{Aoki}},
  \bibinfo{author}{\bibfnamefont{M.}~\bibnamefont{Tokita}},
  \bibinfo{author}{\bibfnamefont{H.}~\bibnamefont{Honjo}},
  \bibinfo{author}{\bibfnamefont{H.~J.} \bibnamefont{Barraza}},
  \bibnamefont{and}
  \bibinfo{author}{\bibfnamefont{H.}~\bibnamefont{Katsuragi}},
  \bibinfo{note}{unpublished}.

\bibitem[{\citenamefont{Richard and Qu{\'e}r{\'e}}(2000)}]{Richard2000}
\bibinfo{author}{\bibfnamefont{D.}~\bibnamefont{Richard}} \bibnamefont{and}
  \bibinfo{author}{\bibfnamefont{D.}~\bibnamefont{Qu{\'e}r{\'e}}},
  \bibinfo{journal}{Europhys. Lett.} \textbf{\bibinfo{volume}{50}},
  \bibinfo{pages}{769} (\bibinfo{year}{2000}).

\bibitem[{\citenamefont{Katsuragi and Durian}()}]{Katsuragi2010}
\bibinfo{author}{\bibfnamefont{H.}~\bibnamefont{Katsuragi}} \bibnamefont{and}
  \bibinfo{author}{\bibfnamefont{D.~J.} \bibnamefont{Durian}},
  \bibinfo{note}{unpublished}.

\end{thebibliography}

\end{document}